\documentclass[letterpaper,aps,prd,10pt,showpacs,twocolumn,eqsecnum,superscriptaddress]{revtex4}

\usepackage{amssymb}
\usepackage{amsmath}
\usepackage{epsfig}
\usepackage{graphicx}
\usepackage{latexsym}
\usepackage{bm,latexsym}
\usepackage{mathrsfs}
\usepackage{appendix}

\begin{document}


\title{Gravitational waves from scalar field accretion}
\author{Dar\'{\i}o N\'u\~nez} \email{nunez@nucleares.unam.mx} \affiliation{Instituto de Ciencias Nucleares, Universidad
Nacional Aut\'onoma de M\'exico, Apdo. 70-543, CU, 04510 M\'exico,
D.F., M\'exico}
\author{Juan Carlos Degollado} \email{jcdegollado@ciencias.unam.mx} \affiliation{Instituto de Astronom\'ia, Universidad
Nacional Aut\'onoma de M\'exico, Apdo. 70-264, CU, 04510 M\'exico,
D.F., M\'exico}
\author{Claudia Moreno} \email{claudia.moreno@cucei.udg.mx} \affiliation{Departamento de Matem\'atica y Postgrado en F\'isica\\
Centro Universitario de Ciencias Exactas e Ingenier\'ia, Universidad de Guadalajara\\
Av. Revoluci\'on 1500, Colonia Ol\'impica C.P. 44430, Guadalajara, Jalisco, M\'exico}


\date{\today}

\label{firstpage}


\begin{abstract}
Our aim in this work is to outline some physical consequences of
the interaction between black holes and scalar field halos
in terms of gravitational waves. In doing so,
the black hole is taken as a static and spherically symmetric gravitational source, {\it i. e.}
the Schwarzschild black hole, and we work within the test field approximation, considering that the
scalar field lives in the curved space-time outside the black
hole.
We focused on the emission of gravitational waves when the black hole
is perturbed by the surrounding scalar field matter. The symmetries of the spacetime and
the simplicity of the matter source allow, by means of a spherical harmonic decomposition,
to study the problem by means of a one dimensional description.
Some properties of such gravitational waves are discussed as a function of the parameters of the
infalling scalar field, and allow us to make the conjecture that the gravitational waves carry
information on the type of matter that generated them.


\end{abstract}
\maketitle


\section{Introduction}
\label{sec:introduction}

The scalar field is a current research subject in a large variety of areas.
In particular, in Astrophysics, the scalar field has played a central role in developing
alternative theories to the general relativity theory \cite{Brans:1961sx}; and more particular to
our aim, in developing the possible existence of stable compact objects, the boson stars
\cite{boson} or the oscillatons \cite{Seidel:1993zk,Alcubierre:2003sx}. Furthermore, the scalar
field is a solid candidate for describing the nature of dark matter
\cite{Matos:2000ng,Matos:2000ki}.

The fact that the scalar field interacts negligibly with barionic matter, makes it difficult to
detect scalar objects and, also, to test the viability of the scalar field as dark matter candidate,
a feature common to any dark matter model. Given this problem, one could wonder about the
possibility, at least in principle, that the gravitational waves associated with the dynamics of a
scalar field, either by collision of scalar objects, or by accretion of a scalar field unto a black
hole, would carry information that identified them as generated by such scalar field dynamics.
Specifically, when a black hole is perturbed by the infall of matter, do the gravitational waves
generated carry some information on the type of matter which generated them?

The theory of scattering of scalar pulses by black holes is very well established, (see, for
instance, \cite{Burko:2004jn,Koyama:2001ee}) however, the response of the black hole to such
accretion has not been
developed. In this paper we fill this gap and, by means of perturbation theory in the frame of
Newman-Penrose formalism, we study the problem of accretion of a scalar field onto a non rotating black
hole and the corresponding gravitational wave generation.

The determination of the sources of gravitational waves which, in turn, could be
observed by the gravitational wave detectors, has been the central problem in
Numerical Relativity Astrophysics. The strongest candidates for the current
detectors are the gravitating binary systems, involving highly non-linear
evolutions of either black holes or neutron stars. The emision of gravitational waves
due to the collision of a Boson Stars binary system has also
been studied, \cite{Palenzuela:2006wp,Palenzuela:2007dm}, showing a different, richer, behavior of
the gravitational profile than that corresponding to the black hole collision. These results show
that indeed, the gravitational waves carry some information about the sources which generated them.
The case of a black hole - boson star collision has, to our knowledge, not been discussed. We
hope that the results presented in this manuscript will encourage the research along this line.

We want to mention also the phenomenon known as gravitational wave burst with memory, proposed by 
Braginski and K. Thorne and by D. Christodoulou \cite{Braginski:1987,Christodoulou:1991cr}, see
also \cite{Thorne92}, and \cite{Favata:2009ii} for a recent development. This phenomenon describes
how a distribution
of test particles, once a gravitational wave passes, does not go back to the original distribution, 
these particles have "a memory" that the wave passed. The phenomenon appears also by the fact that
at late time and early time values of at least one 
of the gravitational wave polarizations differ from zero \cite{Favata:2010zu}.

The gravitational wave with memory could be enriched with the results presented in this manuscript.
As the distribution
of test particles has memory that the gravitational wave passed, it would also have an imprint of
the tail
behavior. On the other hand, the tail is a late time manifestation of the gravitational wave, (it is
the backscattering of the gravitational wave due to the spacetime curvature), and is a non linear
effect. Thus, it could be related to the Christodoulou memory (non linear regime), as long as the
memory is sensitive to the entire past history of the process that generates the gravitational wave.
Thus, a detail study  of the tails will
help in obtaining an accurate description of the memory, by means of an analysis similar to the one
performed in \cite{Favata:2009ii}.

As we will see, the gravitational wave generated by the in-fall of the scalar field has some
unique features that differentiate them from those generated by other means, as binary black hole
collision or by the in-fall of dust \cite{JC09}. Indeed, it will be seen that the ring-down part of
the gravitational wave has the same values of the quasi normal frequencies as those obtained in the
case of a binary black hole collision. However, the late time behavior of the gravitational wave
generated by the scalar field in-fall, does differ from the late time behavior of the gravitational
waves generated by other means. Thus, our analysis allow us to conjecture that, by means of the late
time behavior of the gravitational waves, it is possible to infer some properties of their source,
and, in principle, infer the nature of the dark matter halo by means of its imprint in the
gravitational emission generated by accretion unto the central super massive black hole.

Also, as long as it has been shown that the gravitational wave generated by the
perturbation of a thin dust shell falling unto a black hole, has the same features as the one
generated by the final stage of a binary black hole collision \cite{JC09}.
We conjecture that our results give a hint on the profile of gravitational waves formed by the
collision of a black hole with a scalar compact object such a boson star.

Our general line of work in the present manuscript consists in performing a decomposition of the
scalar field in spherical harmonics. Furthermore, by using of the properties of the operators
acting on the scalar field stress energy tensor, we are able to deal with non spherical
distributions within a radial description. We solve numerically the evolution for
different initial distributions of the scalar field and obtain the gravitational wave produced.

The paper is organized as follows. In Section \ref{sec:pert_eq} we discuss the main properties in
the derivation of the perturbation equation for a Schwarzschild space time with arbitrary sources.
In Section \ref{sec:scalarfield} we discuss the
properties of the scalar field and show how the Klein Gordon equation can be separated in
radial-temporal and angular parts, resulting in a radial-temporal equation for each mode.
In Section \ref{sec:numex} we present some results obtained by solving numerically the equation
for the gravitational perturbation when a source is modeled by a scalar field. We
analyze the infall of one Gaussian pulse of scalar field and the gravitational response
due to the variation in the width of such pulse. Then, we study the effect of
the coupling  of the different scalar modes in the gravitational signal. Finally in Section
\ref{sec:dis} we present a discussion of the results.


\section{Perturbation equation}
\label{sec:pert_eq}

The line element of the Schwarzschild black hole,
in the ingoing Kerr-Schild coordinates is given by:
\small{
\begin{equation}
ds^2=-\left(1-\frac{2 M}{r} \right)\,dt^2 + \frac{4 M}{r}\,dt\,dr + \left(1+ \frac{2 M}{r}
\right)\,dr^2 + r^2\,d\Omega^2.
\label{eq:lel_KS}
\end{equation}
}
In order to set up the Newman Penrose formalism we define the null tetrad as:
\begin{eqnarray}
&l^\mu=\frac12\,\left(1+\frac{2 M}{r},1-\frac{2 M}{r},0,0\right); \hspace{1cm}
k^\mu=\left(1,-1,0,0\right);& \nonumber \\
&\hspace{1cm} m^\mu=\frac1{\sqrt{2}\,r}\left(0,0,1,i\,\csc\theta\right). \label{eq:nullv}&
\end{eqnarray}
In \cite{Zeng08a,JC09} was derived in detail the equation for the perturbed Weyl scalar ${\Psi_4}$,
that is associated to gravitational radiation. We consider at this point the case when
the perturbation is produced by an stress energy tensor $T_{\mu\nu}$,
\begin{equation}
\left[{\square}^{\Phi}_{tr} + \square^{\Phi}_{\theta \phi} \right]\,\Phi=16\,\pi\,r^3\,{T},
\label{eq:pertPhi}
\end{equation}
where we have defined the function $\Phi=r\,\,{\Psi_4}$, which is expected to
have a constant behavior in the regions far from the black hole due to the peeling theorem (see
\cite{Bondi62,Sachs62}), and where we have defined the operators:
\begin{eqnarray}
&{\square}^{\Phi}_{tr}=-\left(r^2+2\,M\,r\right)\,\frac{\partial^2}{\partial t^2}
 +  \left(r^2-2\,M\,r\right)\,\frac{\partial^2}{\partial r^2} + &\label{op:Pert1rt} \\
&4\,M\,r\,\,\frac{\partial^2}{\partial t \partial r} + 2\,\left(2\,r+
M\right)\,\frac{\partial}{\partial t} + 2\,\left(2\,r -M\right)\,\frac{\partial}{\partial r} +
2\frac{M}r, \nonumber & \\
&\square^{\Phi}_{\theta\varphi}= \frac{\partial^2}{\partial \theta^2} +
\frac1{\sin^2\theta}\,\frac{\partial^2}{\partial \varphi^2}
+ \cot\theta\,\frac{\partial}{\partial \theta} -
4\,i\,\frac{\cos\theta}{\sin^2\theta}\,\frac{\partial}{\partial \varphi} - & \nonumber \\
&2\,\frac{1+\cos^2\theta}{\sin^2\theta}, &\label{op:Pertthph}
\end{eqnarray}
and the source term $T$:
\begin{equation}
T=\left[{\cal{{\hat T}}}^{k\,k}\,k^\mu\,k^\nu + {\cal{{\hat T}}}^{k\,m^*}\,k^\mu\,{m^*}^\nu
+ {\cal{{\hat T}}}^{m^*\,m^*}\,{m^*}^\mu\,{m^*}^\nu\right]\,T_{\mu\nu}, \label{eq:T4}
\end{equation}
with the operators described as:
\begin{eqnarray}
{\cal{{\hat T}}}^{k\,k}&=&-\frac1{2\,r^2}\,{\bar\eth}_{-1}\,{\bar\eth}_0, \nonumber\\
{\cal{{\hat T}}}^{k\,m^*}&=&-\frac{\sqrt{2}}{r}\,\left(\frac{\partial}{\partial
t}-\frac{\partial}{\partial r} -\frac3r\right){\bar\eth}_{-1}, \label{ops:t}\\
{\cal{{\hat T}}}^{m^*\,m^*}&=&-\frac{\partial^2}{\partial t^2} + 2\,\frac{\partial^2}{\partial t
\partial r} - \frac{\partial^2}{\partial r^2} + \frac6r\left(\frac{\partial}{\partial
t}-\frac{\partial}{\partial
r}\right)  - \frac4{r^2}. \nonumber
\end{eqnarray}
Where we have used the ``eth'' and ``eth-bar'' \cite{Newm66,Gold67}:
\begin{eqnarray}
&\eth_s=-\left(\frac{\partial}{\partial \theta} + i\,\csc\theta\,\frac{\partial}{\partial \varphi}
- s\,\cot\theta\right)\equiv \eth_0 + s\,\cot\theta,& \nonumber \\
&&\label{ops:eth}\\
&{\bar\eth}_s=-\left(\frac{\partial}{\partial \theta} - i\,\csc\theta\,\frac{\partial}{\partial
\varphi}
+ s\,\cot\theta\right)\equiv {\bar\eth}_0 - s\,\cot\theta,& \nonumber
\end{eqnarray}
which act on the spin weighted spherical harmonics ${Y_s}^{l,m}(\theta,\varphi)$, raising and lowering
the spin weight as follows:
\begin{eqnarray}
\eth_s\,{Y_s}^{l,m}&=&\sqrt{\left(l-s\right)\,\left(l+s+1\right)}\,{Y_{s+1}}^{l,m}, \label{eth_Y} \\
{\bar \eth}_s\,{Y_s}^{l,m}&=&-\sqrt{\left(l+s\right)\,\left(l-s+1\right)}\,{Y_{s-1}}^{l,m}.
\label{beth_Y}
\end{eqnarray}
Eq.~(\ref{op:Pertthph}), can be expressed in terms of the
eth operators:
\begin{equation}
\square^{\Phi}_{\theta\varphi}={\bar\eth}_{-1}\,\eth_{-2},
\end{equation}
and it can be shown that the spin weighted spherical harmonics, with spin $-2$, are eigenfunctions
of
such angular operator:
\begin{equation}
\square^{\Phi}_{\theta\varphi}\,{Y_{-2}}^{l,m}=-\left(l-1\right)\,\left(l+2\right)\,{Y_{-2}}^{l,m}~.
\label{eq:Ym2}
\end{equation}
We expanded the gravitational perturbation function in terms of these spin weighted spherical
harmonics
\begin{equation}
\Phi=\sum\limits_{lm}\,\Phi_{l,m}(t,r)\,{Y_{-2}}^{l,m}(\theta,\phi)~,
\label{def:Phi_lm}
\end{equation}
and the perturbation equation, Eq.~(\ref{eq:pertPhi}) becomes:
\begin{equation}
\sum\limits_{lm}\,{Y_{-2}}^{l,m}\,\left[{\square}^{\Phi}_{tr} -\left(l-1\right)\,\left(l+2\right)
\right]\,\Phi_{l,m}=16\,\pi\,r^3\,{T}.
\label{eq:pertPhi1}
\end{equation}
Finally, using the orthonormal properties of the spin weighted spherical harmonics
\begin{equation}
\oint\,\sum\limits_{lm}\,{\bar{Y}_{-2}}^{l'',m''}\,{Y_{-2}}^{l,m}\,d\Omega=\delta_{l,l''}\,\delta_{m
,m''},
\end{equation}
we get rid of the sum over the modes in the left hand side and get a term involving
only the radial-temporal part:
\begin{equation}
\left[{\square}^{\Phi}_{tr} -\left(l''-1\right)\,\left(l''+2\right)
\right]\,\Phi_{l'',m''}=16\,\pi\,r^3\,S,
\label{eq:pertPhi1b}
\end{equation}
where $S=\oint\,{\bar{Y}_{-2}}^{l'',m''}\,{T}\,d\Omega$.
This is the final expression needed in order to describe the gravitational waves. Indeed, the
angular dependence of these waves is given by the spin weighted spherical harmonics, and
this is the equation satisfied by the radial and temporal dependence of each mode of the
gravitation wave, given an specific stress energy tensor, describing the in-fall of the matter
generating the perturbation.

In the present work, we will consider the stress energy tensor of a scalar field, as the source for
the gravitational perturbation. In the next section we describe how to deal with such source.


\section{Scalar field}
\label{sec:scalarfield}

Let us consider the matter source to be described by a scalar field $\chi$, whose stress energy tensor is:
\begin{equation}
 T_{\mu\nu}=\chi_{,\mu}\chi_{,\nu} -
\frac{1}{2}g_{\mu\nu}( \chi^{,\alpha}\chi_{,\alpha}+V(\chi)  )
. \label{eq:tmunu}
\end{equation}
We are neglecting the gravitational field generated by the scalar field
itself, so the dynamics of the scalar field are dictated by the Klein-Gordon equation in the
Schwarzschild
background, Eq.~(\ref{eq:lel_KS}). The equation of motion takes the form:
\begin{equation}
\frac{1}{r^3}\,\left[\square^{\rm sc}_{tr} +
\square^{\rm sc}_{\theta\varphi}\right]\,\left(r\,\chi\right) -
\,\frac{d\,V(\chi)}{d\,\chi}=0,
\label{eq:KG1}
\end{equation}
where we have defined the operators:
\begin{eqnarray}
&\square^{\rm sc}_{tr}=-\left(r^2 + 2\,M\,r\right)\,\frac{\partial^2}{\partial t^2} +
\left(r^2 - 2\,M\,r\right)\,\frac{\partial^2}{\partial r^2} + & \nonumber
\\ & 4\,M\,r\,\frac{\partial^2}{\partial t
\partial r} - 2\,M\,\left(\frac{\partial}{\partial t}
- \frac{\partial}{\partial r}\right) - 2\,\frac{M}{r},&
\label{op:Pertrt_esc} \\
& \square^{\rm sc}_{\theta\varphi}= \frac{\partial^2}{\partial \theta^2}
+ \frac1{\sin^2\theta}\,\frac{\partial^2}{\partial \varphi^2} +
\cot\theta\,\frac{\partial}{\partial \theta}. & \label{op:Pertthph_esc}
\end{eqnarray}
Eq. \eqref{op:Pertrt_esc} is for $r\,\chi$, where again we are taking into
account
the expected behavior of the scalar field in the region far from the black hole,
which goes as $\sim 1/r$.

We will consider a scalar potential of the form $V(\chi)=\frac12  \,\mu^2\,\chi^2$,
where $\mu$ is related to the mass of the scalar field $m_\chi$, by
$\mu=\frac{m_\chi\, c}{\hbar}$,  where $c$ is the speed of light and $\hbar$
the Planck constant. The dynamical equation, Eq.~(\ref{eq:KG1}), takes the form:
\begin{equation}
\left[\square^{\rm sc}_{rt} + \square^{\rm sc}_{\theta\varphi} -
\,r^2\, \mu^2\right]\,\left(r\,\chi\right)=0.
\label{eq:KG2}
\end{equation}
Again, as long as the spherical harmonics of spin zero are eigenfunctions of the angular
operator \eqref{op:Pertthph_esc} it follows that:
\begin{equation}
\square^{\rm sc}_{\theta\varphi}\,{Y_{0}}^{l,m}=-l\,\left(l+1\right)\,{Y_{0}}^{l,m}.
\label{eq:eignY}
\end{equation}
Then, we perform the following expansion:
\begin{equation}
\chi=\sum\limits_{lm}\,\chi_{l,m}(t,r)\,{Y_{0}}^{l,m}(\theta, \varphi). \label{eq:chi_lm}
\end{equation}
Replacing this expression in the Klein Gordon equation, together with Eq.~(\ref{eq:eignY}),
the dynamical equation for the scalar field becomes
\begin{equation}
\sum\limits_{lm}\,{Y_{0}}^{l,m}\,\left[\square^{\rm sc}_{tr} - l\,\left(l+1\right) -
r^2\,\mu^2\right]\,\left(r\,\chi_{l,m}\right)=0.
\end{equation}
It follows that each mode must satisfy that:
\begin{equation}
\left[\square^{\rm sc}_{tr} - l\,\left(l+1\right) -
r^2\,\mu^2\right]\,\left(r\,\chi_{l,m}\right)=0.
\label{eq:KG3}
\end{equation}
The radial and temporal dependence are determined by this last $1+1$ equation. In this way, given an
initial condition on the scalar field surrounding the black hole, we determine the evolution of this
field by means of Eq.~(\ref{eq:KG3}).
With this field, we construct the corresponding stress energy tensor, and
apply the needed operators on such tensor, in order to determine the final source term $S$ in the
gravitational perturbation equation, Eq.~\eqref{eq:pertPhi1b}.
The explicit derivation and construction of the source term $S(\chi_{lm})$ is as follows:
starting from the scalar field stress energy tensor Eq.~\eqref{eq:tmunu},
we perform the corresponding projections along the null vectors needed in Eq.~\eqref{eq:T4}:
\begin{eqnarray}
&k^\mu\,k^\nu\,T_{\mu\nu}=\left(k^\mu\,\chi_\mu\right)^2=\left(\sum\limits_{lm}\,A_{l,m}\,{Y_{0}}^
{l,m}\right)^2,& \label{proy:kk}\\
&k^\mu\,{m^*}^\nu\,T_{\mu\nu}=\left(k^\mu\,\chi_\mu\right)\,
\left({m^*}^\mu\,\chi_\mu\right)=& \label{proy:kmc} \\
&-\frac1{\sqrt{2}\,r}\,\left(\sum\limits_{lm}\,A_{l,m}\,{Y_{0}}^{l, m}\right)\,
\left(\sum\limits_{lm}\,\sqrt {l\,\left(l+1\right)}\,\chi_{l,m}\,{Y_{-1} }^{l,m}\right),&
\nonumber \\
&{m^*}^\mu\,{m^*}^\nu\,T_{\mu\nu}=\left({m^*}^\mu\,\chi_\mu\right)^2=&\nonumber \\
&\frac1{2\,r^2}\,\left(\sum\limits_{lm}\,\sqrt
{l\,\left(l+1\right)}\,\chi_{l,m}\,{Y_{-1} }^{l,m}\right)^2,& \label{proy:mcmc}
\end{eqnarray}
also we have taken into account the form of the null vectors, Eq.~(\ref{eq:nullv}) and the fact
that
\begin{equation}
{m^*}^\mu\,\partial_{\mu}=\frac1{\sqrt{2}\,r}\,\left(\partial_\theta
-\frac{i}{\sin\theta}\partial_\varphi\right)=-\frac1{\sqrt{2}\,r}\,\bar{\eth}_0,
\end{equation}
and we have defined the quantities,
\begin{equation}
A_{l,m}=\partial_t\,\chi_{l,m} - \partial_r\,\chi_{l,m}.
\end{equation}
Notice that the action of the {\it eth} operator on the spherical harmonics are given in
Eq.~(\ref{ops:eth}) for $s=0$. Applying it into the scalar field, we get:
\begin{eqnarray}
&{m^*}^\mu\,\chi_\mu=-\frac1{\sqrt{2}\,r}\,\bar{\eth}_0\,
\sum\limits_{lm}\,\chi_{l,m}\,{Y_{0}}^{l,m}=& \nonumber
\\ &\frac1{\sqrt{2}\,r}\,\sum\limits_{lm}\sqrt{l\,(l+1)}\,
\chi_{l,m}\,{Y_{-1}}^{l,m}.&
\end{eqnarray}
The action of the {\it eth} operators on pairs of
spherical harmonic is the following:
\begin{eqnarray}
&\bar{\eth}_{-1}\,{Y_{0}}^{l_1,m_1}\,{Y_{-1}}^{l_2,m_2}= -\sqrt{(l_2-1)\,(
l_2+2)}\,{Y_{0}}^{l_1,m_1}\,{Y_{-2}}^{l_2,m_2} & \nonumber\\
& -\sqrt{l\,(l_1+1)}\,{Y_{-1}}^{l_1,m_1}\,{Y_{-1}}^{l_2,m_2}.&
\\
&\bar{\eth}_{-1}\,\bar{\eth}_0\,{Y_{0}}^{l_1,m_1}\,{Y_{0}}^{l_2,m_2} =&\nonumber\\
& \sqrt{(l_1-1)\,l_1\,(l_1+1)\,
(l_1+2)}\,{Y_{0}}^{l_2,m_2}\,{Y_{-2}}^{l_1,m_1}  & \nonumber \\
&+\sqrt{(l_2-1)\,l_2\,(l_2+1)\,(l_2+2)}\,{Y_{0}}^{l_1,m_1}\,{Y_{-2}}^{l_2,
m_2}  &\nonumber \\
&
+2\,\sqrt{l_1\,(l_1+1)\,l_2\,(l_2+1)}\,{Y_{-1}}^{l_1,m_1}\,{Y_{-1}}^{l_2,m_2}. &
\end{eqnarray}
Applying the radial-temporal operators on $\chi_{l,m}(t,r)$, we finally obtain the
expression for the source term for the gravitational perturbation due to the scalar field.

In order to have a clearer understanding for the process in a particular example, we
present explicitly the case when the scalar field is expanded up to the $l=2$
harmonic mode number, (there is a sum over $m_i$) with $|m_i| \leq l_i$, $(i=0,1,2)$.
\begin{equation}
\chi=\chi_{0,0}\,{Y_{0}}^{0,0} + \chi_{1,m_1}\,{Y_{0}}^{1,m_1} + \chi_{2,m_2}\,{Y_{0}}^{2,m_2}.
\label{eq:chi_3m}
\end{equation}
The derivatives along the null directions are:
\begin{eqnarray}
k^\mu\,\chi_\mu&=& A_{0,0}\,{Y_{0}}^{0,0}+A_{1,m_1}\,{Y_{0}}^{1,m_1} +
A_{2,m_2}\,{Y_{0}}^{2,m_2},\nonumber
\\
&&
\\
{m^*}^\mu\,\chi_\mu&=&\frac1{r}\,\left(\chi_{1,m_1}\,{Y_{-1}}^{1,m_1} +
\sqrt{3}\,\chi_{2,m_2}\,{Y_{-1}}^{2,m_2}\right),\nonumber
\end{eqnarray}
the corresponding projections on the stress energy tensor are
\begin{eqnarray}
&k^\mu\,k^\nu\,T_{\mu\nu}=\left(A_{0,0}\,{Y_{0}}^{0,0}+A_{1,m_1}\,{Y_{0}}^{1,m_1} +
A_{2,m_2}\,{Y_{0}}^{2,m_2}\right)^2,&
\nonumber \\
&k^\mu\,{m^*}^\nu\,T_{\mu\nu}=\frac{1}{r}\,\left(A_{0,0}\,{Y_{0}}^{0,0}+A_{1,m_1}\,{Y_{0}}^{1,m_1} +
A_{2,m_2}\,{Y_{0}}^{2,m_2}\right)\times &
\nonumber \\
&\left(\chi_{1,m_1}\,{Y_{-1}}^{1,m_1} +
\sqrt{3}\,\chi_{2,m_2}\,{Y_{-1}}^{2,m_2}\right),&  \\
&{m^*}^\mu\,{m^*}^\nu\,T_{\mu\nu}=\frac1{r^2}\,\left(\chi_{1,m_1}\,{Y_{-1}}^{1,m_1} +
\sqrt{3}\,\chi_{2,m_2}\,{Y_{-1}}^{2,m_2}\right)^2.&\nonumber
\end{eqnarray}
The expression for the source term in this particular case is
\begin{eqnarray}
&T=\frac{2}{r^2}\,\left({\cal{A}}_0\,{Y_{0}}^{0,0}\,{Y_{-2}}^{2,m_2}
+ {\cal{A}}_1\,{Y_{0}}^{1,m_1}\,{Y_{-2}}^{2,m_2} + \right. & \nonumber \\
& \left. {\cal{A}}_2\,{Y_{0}}^{2,m_2}\,{Y_{-2}}^{2,m_2} + {\cal{A}}_3\,\left({Y_{-1}}^{1,m_1}\right)^2
+ {\cal{A}}_4\,\left({Y_{-1}}^{2,m_2}\right)^2
+ \right. & \nonumber \\
& \left.  {\cal{A}}_5\,\left( {Y_{-1}}^{1,m_1}\,{Y_{-1}}^{2,m_2} \right) \right. ,  &  \label{eq:Tf}
\end{eqnarray}
where
\begin{eqnarray}
&{\cal{A}}_0=\sqrt{6}\,\chi_{2,m_2}\,\left(\partial_t\,A_{0,0} -
\partial_r\,A_{0,0} - 2\,\frac{A_{0,0}}{r}\right), &
\nonumber \\
&{\cal{A}}_1=\sqrt{6}\,\chi_{2,m_2}\,\left(\partial_t\,A_{1,m_1} -
\partial_r\,A_{1,m_1} - 2\,\frac{A_{1,m_1}}{r}\right), &
\nonumber \\
&{\cal{A}}_2=\sqrt{6}\,\chi_{2,m_2}\,\left(\partial_t\,A_{2,m_2} -
\partial_r\,A_{2,m_2} - 2\,\frac{A_{2,m_2}}{r}\right), \nonumber & \\
&{\cal{A}}_3=-\left(\left({A_{1,m_1}}\right)^2 - \left(\frac{\chi_{1,m_1}}{r}\right)^2\right), \nonumber
&
\\
&{\cal{A}}_4=-3\,\left(\left({A_{2,m_2}}\right)^2 - \left(\frac{\chi_{2,m_2}}{r}\right)^2\right),&
\label{coef:T} \\
&{\cal{A}}_5=-2\sqrt{3}\,\left( A_{1,m_1}A_{2,m_2} -\frac{\chi_{1,m_2}\chi_{2,m_2}}{r^{2}}\right).\nonumber&
\end{eqnarray}

Finally, we multiply by
the conjugate of ${{Y}_{-2}}^{l_3,m_3}$, and perform the integration over the solid angles, that is,
$S=\oint\,{\bar{Y}_{-2}}^{l_3,m_3}\,{T}\,d\Omega$.
It can be seen that this implies the integration over the solid angle of the product of three spin
weighted spherical
harmonics: $\oint\,{Y_{2}}^{l'',m''}\,{Y_{s_2}}^{l_2,m_2}\,{Y_{s_3}}^{l_3,m_3}\,d\Omega$. These
integrals are well known in terms of the Wigner matrices, see for example \cite{Ruiz:2007rs},
and have a closed expression.

Using these expressions, the evaluation of the angular matrices appearing in the expression for the
source term $T$, is straightforward.

For the case when the scalar field is described by the
monopole, dipole and quadrupole components, Eq.~(\ref{eq:chi_3m}), the source term is:
\begin{equation}
S(\chi_{l,m})=\frac{2}{r^2}\,\sum\limits_{i=0}^5\,C_i\,{\cal{A}}_i(\chi_{l,m}),
\quad(l=1,2,3) \label{eq:S}
\end{equation}
where the functions ${\cal{A}}_i$ are given in Eq.~(\ref{coef:T}), and the coefficients $C_i$
are the corresponding integrals of three spherical harmonics determined by
the expressions given by the Wigner matrices.

There are some remarkable facts to notice right away in the source term. First, notice that the
monopolar term of the scalar field, Eq.~(\ref{eq:chi_3m}), does have an influence on the
gravitational response, by means of its coupling with quadrupole mode.
Since the first term of Eq.~(\ref{eq:Tf}) in general is different from zero.
Also, it can be seen in
the source term that the dipole mode of the scalar field, generates a gravitational response
without the need of coupling with other modes. Indeed, the dipole term, as any other mode of
the scalar field, evolves by itself, without coupling, following the evolution determined by the
Klein-Gordon equation, Eq.~(\ref{eq:KG3}), with $l=1$. The source for the gravitational wave,
however, not only implies a coupling among the modes, but the action of the operators on the
stress energy tensor. Indeed, from Eqs. (\ref{proy:kk})-(\ref{proy:mcmc}), we see that the projections
of this tensor on the null tetrad, generates terms with spherical harmonics of weight $-1$; the
action of the operators given in Eqs.~(\ref{ops:t}), further changes the spin weight of the original
modes, as well as the coupling among them. This explains how an initial dipole mode awakes a
gravitational response which is quadrupole, even without coupling.

In the next section, we will present several cases of the quadrupole gravitational radiation,
generated by different initial configurations of the scalar field.


\section{Scalar field accretion and gravitational response}
\label{sec:numex}
In order to solve numerically the Eqs. \eqref{eq:pertPhi1b}, we define the auxiliary variables:
\begin{equation}
\Psi_{l,m}^{(g)}:=\partial_r \Phi_{l,m}\quad {\rm and}\quad
\Pi_{l,m}^{(g)}:=\frac{r+2\,M}r\,\partial_t\,
\Phi_{l,m} -2\,\frac{M}r\,\Psi_{l,m}^{(g)}~.
 \label{eq:Pig}
\end{equation}
Then the following first order linear system of evolution equations is obtained for the
gravitational wave:\\
\begin{eqnarray}
&&\partial_t\,\Phi_{l,m}=\frac{1}{r+2\,M}\left(r\,\Pi_{l,m}^{(g)} + 2\,M\,\psi_{l,m}^{(g)} \right)~,
\label{eq:evolR}\\
&&\partial_t\,\psi_{l,m}^{(g)}=\partial_r\left(\frac{1}{r+2\,M}\left(r\,\Pi_{l,m}^{(g)} +
2\,M\,\psi_{l,m}^{(g)}
\right)\right), \, \label{eq:evolPsi} \\
&&\partial_t\,\Pi_{l,m}^{(g)}=\frac{1}{r+2\,M}\left(2\,M\,\partial_r\,\Pi_{l,m}^{(g)}
+r\,\partial_r\,\psi_{l,m}^{(g)} \right)  \nonumber \\
&& + \frac{2}{r\,(r+2\,M)^2}\left(\left(2\,r^2+5\,M\,r+4\,M^2\right)\,\Pi_{l,m}^{(g)} + \right. \nonumber
\\
&& \left. \left(r+4\,M\right)\,\left(2\,r+3\,M\right)\psi_{l,m}^{(g)} \right) \nonumber \\
&&+ \left(2\frac{M}{r^3} -
\frac{\left(l-1\right)\,\left(l+2\right)}{r^2}\right)\,\Phi_{l,m} + S(\chi_{l,m}).
\label{eq:evolPig}
\end{eqnarray}
We also perform a first order decomposition to obtain the dynamics of the scalar field.
The equations are quite similar to the gravitational case under the same
first order decomposition
\begin{eqnarray}
\Psi_{l,m}^{(sc)}&:=&\partial_r (r\,\chi_{l,m}), \\ \nonumber
\Pi_{l,m}^{(sc)}&:=&\frac{r+2\,M}r\,\partial_t\,
(r\,\chi_{l,m}) -2\,\frac{M}r\,\Psi_{l,m}^{(sc)}~,
 \label{eq:Pi}
\end{eqnarray}
the temporal derivative satisfies:
\begin{eqnarray}
&\partial_t\,\Pi^{sc}_{l,m}=\frac{1}{r+2\,M}\left(2\,M\,\partial_r\,\Pi_{l,m}^{sc}
+
r\,\partial_r\,\psi_{l,m}^{sc} \right),& \label{eq:evolscalPi} \\
&+ \frac{2}{r\,(r+2\,M)^2}\left(  \psi_{l,m}^{sc}-\Pi_{l,m}^{sc}   \right) +
 \left(\frac{2M}{r^3}-\frac{l(l+1)}{r^2}+\mu^{2}\right)\chi_{l,m}.&
\nonumber
\label{eq:evolPisc}
\end{eqnarray}
We used the method of lines with a third order Runge Kutta time integration and finite
differencing with sixth order accurate stencils. The inner boundary is
chosen to be inside the horizon, since this region is causally
disconnected from the exterior no boundary conditions are required.
The outer boundary is typically located at $r=800M$ (although in some cases we put it
at $r=1600M$), and at the last point of the grid we set up all the incoming modes to zero.
\subsubsection{Dependence of the gravitational signal on the mass of the scalar field }
\label{sec:mass}

\begin{figure}[ht!]
\includegraphics[width=0.47\textwidth]{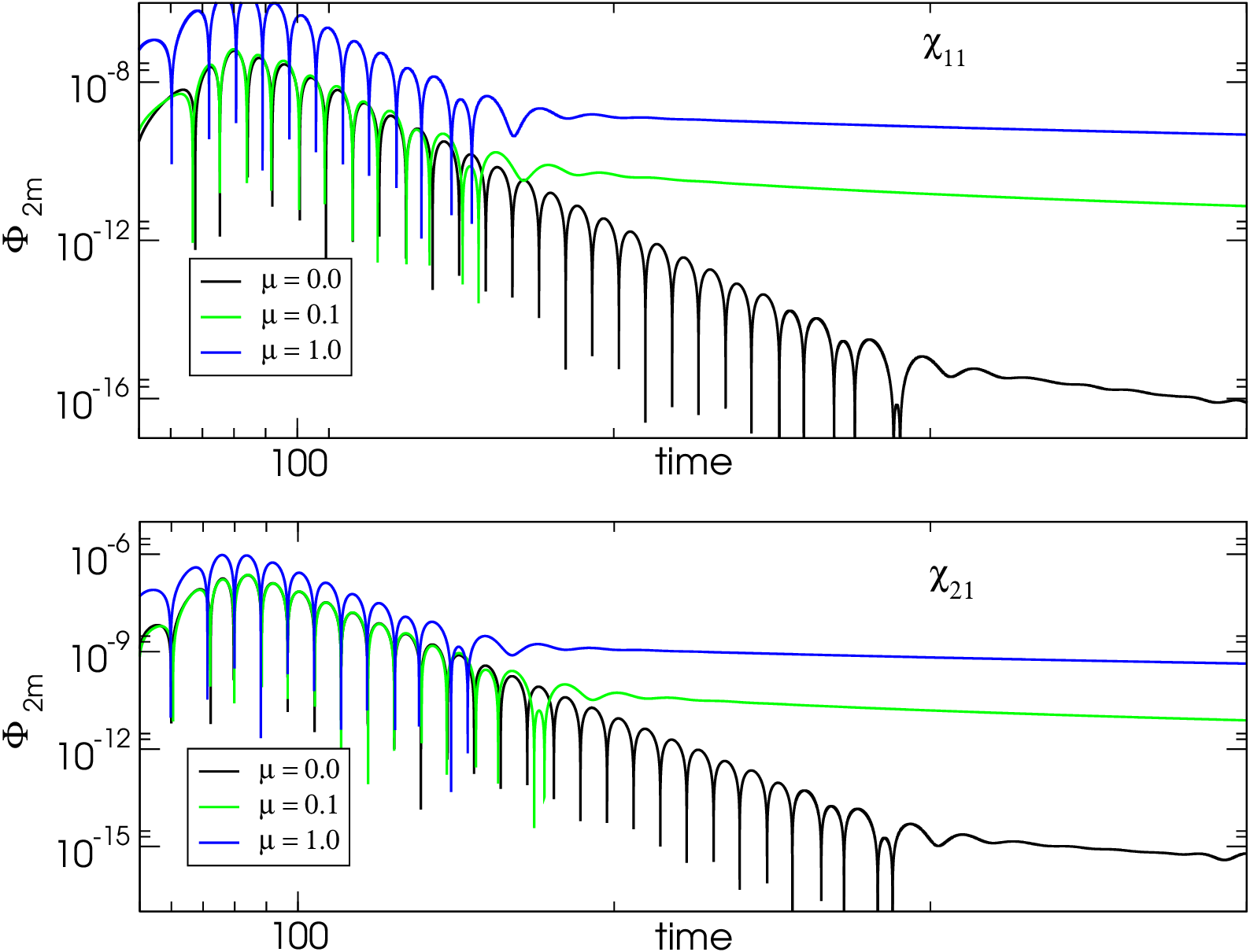}
\caption{As the mass of the scalar field increases, the amplitude of the gravitational
signal produced by the accretion increases. We show the signal measured by an observer
located at 70M. Notice there is no change in the frequency of the signal that corresponds
to different values on the scalar field perturbation. In the panel above the mode was $\chi_{1,1}$,
in the panel below
$\chi_{2,1}$.
}
\label{fig:1}
\end{figure}

As a first approximation, we study the behavior of the gravitational signal when only
one mode of  scalar field falls onto the black hole.
We have used different values for the mass of the scalar field. In \cite{urena},
it is shown that the accretion rate of the scalar field depends strongly on the
values of $M$ and $\mu$. This is due to the relation between the Compton wavelength of the scalar field
and the Schwarzschild radius. If the wavelength is greater
than the size of the black hole, almost all the field will be scattered, on the
other hand if the  wavelength is smaller, the accretion is very likely.
To model the accretion of the scalar field cloud onto the black hole, we start
with the harmonic decomposition, Eq.~\eqref{eq:chi_lm}. In order to have a better understanding
of the process, we consider that the field has a dipole component only {\it i.e.}
the unique coefficient different from zero of the expansion given in Eq.~\eqref{eq:chi_lm} is the
one that multiplies
at $Y_{1,1}$, the coefficient $\chi_{1,1}$. As an initial data we chose a Gaussian perturbation
of the form $\chi_{1,1} = A e^{-(r-r_0)^2/\sigma^{2}}/r$
where $r_0$ corresponds to the center of the Gaussian, $A$ is its amplitude
and $\sigma$ is its width. For a typical run the pulse is centered at $r_0=10M$.
The gravitational mode $R_{2,m}$ on the other hand, is initially set to zero.
With this initial data we are interested in the gravitational
signal produced when the scalar field falls into the black hole.
As expected, the picture is that part of the initial scalar field falls into
the black hole and some of the field is scattered away by the black hole. As a
result, one can expect that the gravitational wave is determined by the properties
of the black hole and, also, by the properties of the scalar field scattered.
We considered an interval of the mass of the scalar field from 0 up to 1 in order to study the
imprint of the scalar field mass on the gravitational signal.
In figure \ref{fig:1} we plot in log scale the gravitational response for this in-falling matter.
One can see that the amplitude increases as the mass increases.
Notice also that the ring down behavior is
shorter as the mass of the scalar field increases, allowing to tail behavior
appear at earlier times.
We repeated  this situation but  with an initial data of the form  $\chi_{2,1} = A
e^{-(r-r_0)^2/\sigma^{2}}/r$.
The result is that for the massless case the amplitude of the signal becomes greater, but the frequency
of the signal is unchanged. The tail behavior is, however, somehow different
and presents an unique feature.
For the massless case the tail of the gravitational wave generated is similar to the standard case
\emph{i.e.} goes like $\sim t^{-7}$. However for the massive case, the tail of the wave generated
goes like $\sim t^{-2.5}$, see figure \ref{fig:1}. This result could be used
as a basis for a deeper study on the tail behavior of gravitational waves in the presence of matter.
As long as it could be used to discriminate the type of matter that generated the gravitational
signal.
In order to get the late time behavior we used a sixth order stencil in the radial derivatives of
the functions. Since the amplitude of the signal in the massless case is very small, we were unable
to follow the signal for long computational times since we reach the roundoff error. However, we are
confident that the signal follows this behavior as long as we acquire the convergence desired. The
final power law was fitted using the interval of time from $200$ to $300$ for the massive case and
$300$ to $400$ for the massless one.

\subsubsection{Dependence on the properties of the scalar field}
\label{sec:properties}
\begin{figure}[ht!]
\includegraphics[width=0.47\textwidth,height=5cm]{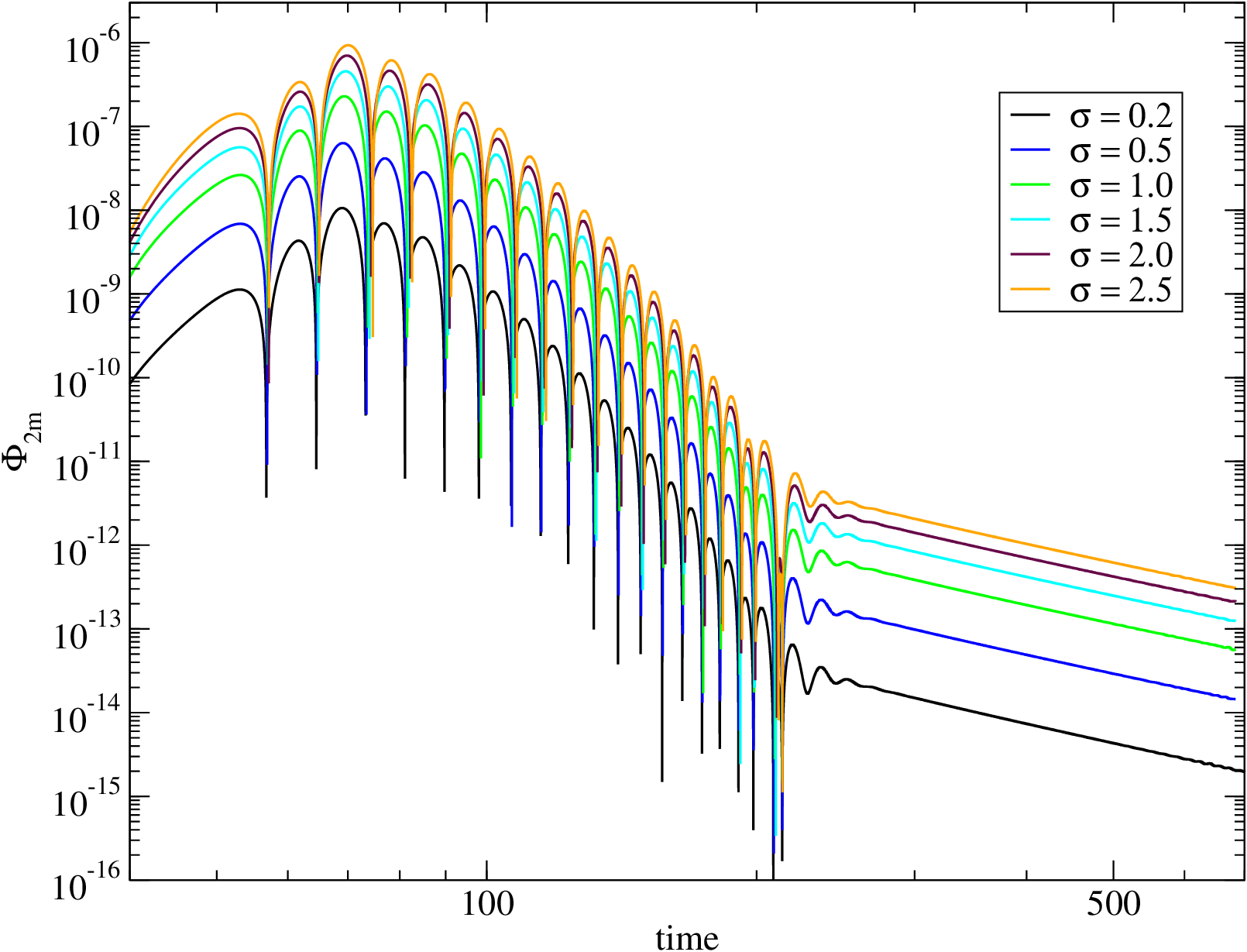}
\caption{It is shown the gravitational signal $\Phi_{2,m}$ produced by the in-falling of a
configuration of the scalar filed composed with the harmonic $Y_{1,1}$ for different values
of $\sigma$. Notice that the difference is only reflected in the amplitude, while the other
properties like the frequency of the ring down or the tail remain the same.}
\label{fig:diffs0}
\end{figure}

A study of the accretion process of dust shells onto a Schwarzschild black hole
was done in references \cite{Font98,JC09}. In these works it was shown that the
properties of the shells play an important role in the gravitational signal produced.
If the shell is compact enough then the corresponding gravitational wave is the one
that corresponds to the quasi normal modes of the black hole. But if the shell is very
spread then the in-falling matter does not produce the typical ring down behavior  of
the black hole. When the in-falling matter is a scalar field there is not such behavior.
The ring down is produced independently of the width values of the initial perturbation
that we considered. The change in compactness is only reflected in the amplitude of
the gravitational signal. In figure \ref{fig:diffs0} it is shown the gravitational
coefficient $\Phi_{2,m}$ when a pulse of scalar field with different widths is accreatted.
We run the code for several values of the $\sigma$ parameter in the initial Gaussian from
$\sigma=0.2$ up to $2.5$. In this figure the mass of the scalar field was $\mu=0.01$ and
the initial amplitude of the perturbation was the same for all the cases.

\begin{figure}[ht!]
\includegraphics[width=0.47\textwidth]{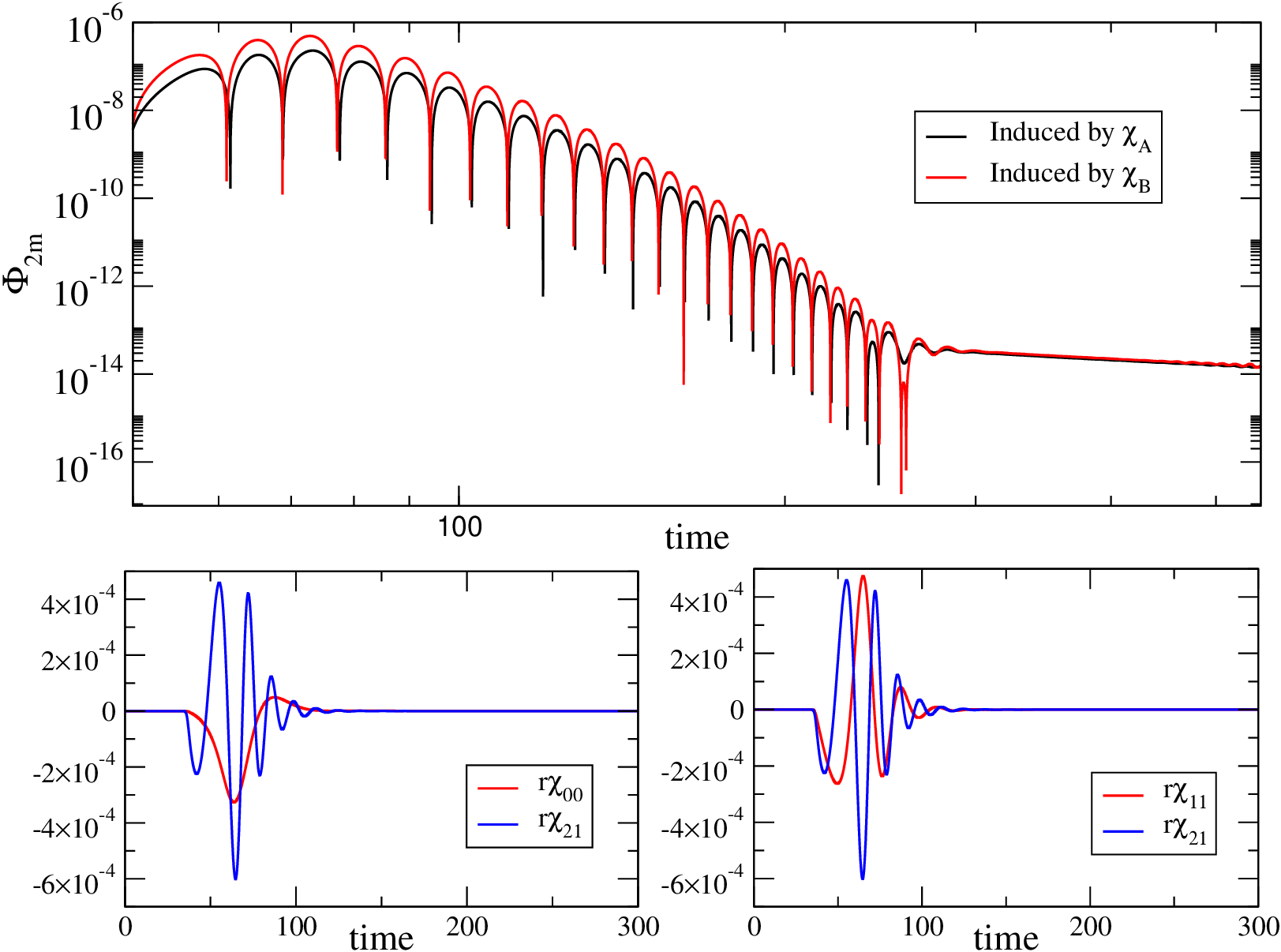}
\caption{In the above panel we superpose (in log scale) the gravitational signal produced by the
in-falling
scalar field when it is composed by two modes
$\chi_{A}=\chi_{0,0}\,{Y_{0}}^{0,0} +  \chi_{2,2}\,{Y_{0}}^{2,2}$ and compare with the case
when the gravitational wave is produced by the in-falling of a couple of different modes,
$\chi_{B}= \chi_{1,1}\,{Y_{0}}^{1,1} + \chi_{2,1}\,{Y_{0}}^{2,1}$. The amplitude of the
former is lower,
however, the frequencies and the late time behavior is the same in both cases. In the bottom panels
we show the signal of the massless scalar field, multiplied by $r$ measured by an observer located at
$r=70M$.}
\label{fig:gravR012}
\end{figure}

As mentioned above, an important remark on the gravitational wave generated by
the scalar field accretion is that the monopolar scalar mode does have an
influence on the gravitational wave, indeed as the zero mode couples to the 2
mode in the sources of the perturbed equation, Eq.~\eqref{eq:pertPhi1}, this
coupling implies that gravitational waves are produced when a spherical
configuration of the scalar field surrounding a black hole, is perturbed. In
the decomposition of the scalar field Eq.~\eqref{eq:chi_3m}, we assume that
the field is formed by the linear superposition of three modes that obey the
Klein-Gordon equation independently. In figure \ref{fig:gravR012} we show the
gravitational signal that is produced when the in-falling scalar field has
monopolar and quadrupole modes. Notice that there is a change in the amplitude
of the signal but the frequencies are the same, ie. the quasi normal modes of
the black hole. Thus, the accretion of the monopolar component of the scalar
field, plays a role in enhancing the gravitational signal.

We performed a similar analysis for the dipole and quadrupole modes.
The results are shown in figure \ref{fig:gravR012}, where it can be
seen that, even though the dynamics of the scalar field is quite
different, the gravitational response is only modified in its amplitude.
Notice also that the tail behavior in this case is unaffected.

\section{Discussion}
\label{sec:dis}

Following the accretion processes described in references
\cite{JC09,Font98,Papadopoulos:2000gb,Nagar:2006eu},
we have studied numerically the gravitational response of a black hole for
several cases of the in-falling scalar field.

We have discussed the importance of in-falling scalar field in the generated gravitational
signal, showing how the ring-down behavior is preserved, unlike the dust shell case in which the
width of the pulse was crucial in order to produce the quasi normal ringing of the black hole.
The scenario in which our model may be important is the one in which an halo of dark matter,
modeled as an scalar field,  surrounds a black hole. Since it is expected that the field falls
into the black hole a gravitational signal may be produced as a result of the the perturbation
onto the black hole.

We studied the properties of the gravitational wave produced by accretion of a test scalar
field. In particular, we were interested on the properties of the field that are reflected
in the gravitational signal. Such
properties are the mass of the boson, the amplitude of the perturbation, and the radial
extension of the field as a bulk. We found that, unlike the case of a dust shell the
compactness of the shell does not affect the ring down behavior of the gravitational wave.
We also found that although the frequency of the scalar field propagating on a Schwarzschild
background is different from the one associated with the gravitational perturbation,
the gravitational signal preserves its natural frequency (the characteristic ring down) no
matter that the scalar signal travels together the gravitational one. In particular we reproduced
the behavior of mass and massless scalar fields in a Schwarzschild space time that has been
studied in detail for both linearized and fully nonlinear evolutions \cite{Burko:2004jn}.

The late time behavior of a massive scalar field in the Schwarzschild space time was studied both
numerically and analytically in the past by \cite{Burko:2004jn} and \cite{Koyama:2001ee}. These
authors showed that the tail behavior of the scalar field goes like $t^{-5/6}\sin(\mu t)$. We found
that the tail behavior of the gravitational signal is in fact modified by the scalar field and, as
mentioned in the introduction, this fact allows us to determine some properties of the source which
generated such gravitational wave.

It is known that the tail behavior of a scalar field depends strongly of the properties 
of the potential. In reference \cite{Hod:2002gb} the author considered a time dependent potential 
and showed that the \emph{tail} behavior actually changes from the standard time independent 
potential.
This phenomena was also considered in references \cite{Chirenti:2011rc} where the authors
considered 
a Vaidya spacetime where the explicit change in the parameters of the black hole is reflected in the
quasi normal frequencies of the scattered field.

The present work gives a solid base to analyze the gravitational waves generated by the scalar
field accretion onto a black hole, and sets us in a track that should be follow on other cases, by
continuing the comparison of gravitational waves generated by different sources in several
configurations, as well as a direct analysis on the dynamical effects of the different
gravitational waves on the actual interferometers. Using the formulation described in
\cite{Nunez:2010ra}, the study presented in this work could be generalized to the case of gravitational
waves from a spinning black hole, generated by the infall of scalar field. This task is currently
under way.


\acknowledgments
The authors are grateful to Miguel Alcubierre, Carlos Palenzuela, and Beatriz Rumbos for
careful reading of our manuscript and useful suggestions. We also 
acknowledges DGAPA-UNAM grant IN115311, SEP-CONACYT grant 49924J and DGAPA-UNAM postoctoral grant.


\bibliographystyle{unsrt}
\bibliography{refs}


\end{document}